\documentclass[11pt]{article}
\usepackage[utf8]{inputenc}

\usepackage{amsmath}
\usepackage{amssymb}
\usepackage{amsthm}
\usepackage{authblk}
\usepackage[mode=buildnew,subpreambles=true]{standalone}

\usepackage{algpseudocode}
\usepackage{algorithm}
\usepackage{tikz}
\usetikzlibrary{positioning}
\usepackage{fullpage}
\usepackage{xcolor}

\usepackage{enumitem}

\usepackage{hyperref}
\usepackage{float}
\title{A Critique of Czerwinski's ``Separation of $\pspace$ and $\exptime$"\thanks{Supported in part by NSF grant 
		CCF-2006496.}}

\author{Ian Clingerman}
\author{Quan Luu}
\affil{Department of Computer Science\\University of Rochester\\Rochester, NY 14627, USA}

\newcommand{\condition}{\,\mid \:}
\newcommand{\naturalnumber}{\ensuremath{{\mathbb{N}}}}
\newcommand{\naturalnumberpositive}{\ensuremath{{\mathbb{N}^+}}}

\newcommand{\p}{\ensuremath{{\rm P}}}

\newcommand{\pspace}{\ensuremath{{\rm PSPACE}}}
\newcommand{\dspace}{\ensuremath{{\rm DSPACE}}}
\newcommand{\dtime}{\ensuremath{{\rm DTIME}}}

\newcommand{\exptime}{\ensuremath{{\rm EXP}}}
\newcommand{\etime}{\ensuremath{{\rm E}}}

\newcommand{\Eset}{\ensuremath{{\mathcal{E}}}}
\newcommand{\Pset}{\ensuremath{{\mathcal{P}}}}
\newcommand{\Enot}{\ensuremath{{\mathcal{E}_0}}}
\newcommand{\Pnot}{\ensuremath{{\mathcal{P}_0}}}

\newtheorem{theorem}{Theorem}
\newtheorem{proposition}{Proposition}
\newtheorem{corollary}{Corollary}
\newtheorem{lemma}{Lemma}

\algblockdefx{Decider}{EndDecider}{\textbf{Decider }}{\textbf{End of Decider}}
\algblockdefx{Def}{EndDef}{\textbf{def }}{}

\date{April 28, 2023}

\begin{document}\sloppy
	
\maketitle

\begin{abstract}
Czerwinski's paper ``Separation of $\pspace$ and $\exptime$''~\cite{cze:t:pspaceexp} claims to prove that $\pspace \neq \exptime$ by showing there is no length-increasing polynomial-time reduction from a given $\exptime$-complete set to a given $\pspace$-complete set. However, in this critique, we show that there are fundamental flaws within the paper's approach and provide a counterexample to one of the paper's theorems, which makes the proposed proof that $\pspace \neq \exptime$ insufficient. 
\end{abstract}
	
\section{Introduction}\label{s:intro}
In this critique, we provide a summary and analysis of Czerwinski's ``Separation of \pspace\ and \exptime''~\cite{cze:t:pspaceexp}. The paper purportedly proves its main claim that $\pspace \neq \exptime$ by 
introducing a novel proof technique with three main theorems: Theorem~\ref{t:1} proposes that
any nontrivial property $\textbf{P}$ of the time and space complexity of an arbitrary Turing machine is undecidable, Theorem~\ref{t:2} claims that if sets $\textbf{A}$ and $\textbf{B}$ are both $\exptime$-complete sets, then they are length-increasing one-one reducible to each other, and Theorem~\ref{t:3} argues that there exists an $\exptime$-complete set $\Enot$ and a $\pspace$-complete set $\Pnot$ such that $\Enot$ is not length-increasing one-one reducible to $\Pnot$.
If correct, Czerwinski's paper would be a major advance in computational complexity theory. In particular, it is well-known that $\pspace \subseteq \exptime$. On the other hand, whether $\pspace$ equals $\exptime$ remains one of the major open questions in theoretical computer science. Furthermore, if Czerwinski's paper truly has introduced a novel proof technique to separate $\pspace$ and $\exptime$, that could potentially serve as the foundation to establish other new separations. 

The approach used in Czerwinski's paper to show the separation of $\exptime$ and $\pspace$ is as follows: Use Corollary~\ref{c:2} of Theorem~\ref{t:1} to prove Theorem~\ref{t:3}, and then by Theorem~\ref{t:2} and Theorem~\ref{t:3}, claim that $\pspace \neq \exptime$. We aim to highlight the flaws in each of these theorems as well as present a counterexample to one of the theorems, thus showing that the paper fails to prove its central claim. 

In Section~\ref{s:prelims}, we provide the 
definitions and notations used throughout the original paper and our critique. Sections~\ref{s:sect1},~\ref{s:sect2},~\ref{s:sect3},~\ref{s:sect4}, and~\ref{s:sect5} each present an overview and a discussion of a section of Czerwinski's paper. Lastly, in Section~\ref{s:conclusion}, we conclude our critique. 
	
\section{Preliminaries}\label{s:prelims}
In this section, we present readers with many standard concepts in complexity theory, as well as frequently used notations in Czerwinski's paper. Interested readers can find equivalent definitions of the concepts in most modern textbooks on complexity~\cite{hem-ogi:b:companion,aro-bar:b:complexity,sip:b:introduction-third-edition}.

Let 
$\naturalnumber = \{0, 1, 2, \ldots\}$ and
let $\naturalnumberpositive = \{1, 2, 3, \ldots\}$. Given a string $w$, let $|w|$ denote the length of $w$. We will let $\epsilon$ denote the empty string. 
For any Turing machine (TM) $M$, let $L(M)$ be the language accepted by $M$. 

We also introduce some notations that Czerwinski uses throughout the paper. 
For a single-tape deterministic Turing machine $M$ on input $x$, 
let $s_M(x)$ be the tape space used by $M$ on input $x$ and let $t_M(x)$ be the time used by $M$ on input $x$. Lastly, a property $\mathbf{P}$ is a set of Turing machines. A Turing machine $M$ is said to have property $\mathbf{P}$ if and only if $M\in\mathbf{P}$. A property $\mathbf{P}$ of the functions $t_M$ and $s_M$ is a set of Turing machines such that for every two Turing machines $M_1$ and $M_2$, if $t_{M_1}(x)=t_{M_2}(x)$ and $s_{M_1}(x)=s_{M_2}(x)$ for all $x\in\Sigma^*$, then either $M_1$ and $M_2$ are both in $\mathbf{P}$ or neither of them are in $\mathbf{P}$. A property $\mathbf{P}$ is nontrivial if and only if $\mathbf{P}\neq\emptyset$ and there exists a Turing machine $M$ such that $M\notin\mathbf{P}$. 

Czerwinski's paper also makes use of the S-m-n~Theorem in a number of its arguments. Since explaining the entire theorem would mean defining many concepts in computability theory that are unnecessary to the purpose of this critique, we suggest interested readers directly consult the book ``Introduction to Metamathematics'' by S.C.~Kleene~\cite{kle:b:metamathematics}.

\subsection{Complexity Classes}

We will now define the complexity classes that will appear throughout the critique. 

$\pspace$ is the class of languages that can be accepted by a deterministic Turing machine using a polynomial amount of space, formally defined as $\pspace = \bigcup_{k \in \naturalnumberpositive} \dspace[n^k]$. 

$\p$ is the class of languages that can be decided by a deterministic Turing machine in polynomial time. Formally, $\p = \bigcup_{k \in \naturalnumberpositive} \dtime[n^k]$.

$\etime$ and $\exptime$ are both classes of languages that can be accepted by a deterministic Turing machine in exponential time. However, they are defined differently and represent two different classes of languages. Formally, $\etime = \bigcup_{c \in \naturalnumberpositive} \dtime [2^{cn}]$ and $\exptime = \bigcup_{c \in \naturalnumberpositive} \dtime [2^{n^c}]$. 
$\etime$, $\exptime$, and $\pspace$ are related in the following ways: (1)~$\etime \subsetneq \exptime$, (2)~$\etime \neq \pspace$~\cite{boo:j:comparing-complexity-classes}, and (3)~$\pspace \subseteq \exptime$. 

An oracle can be thought of as a subroutine for determining membership in a certain set.  
An oracle Turing machine is a modified Turing machine that has the additional capability of querying an oracle.
The actual construction of an oracle is generally unimportant and it can be thought of as a black box. If $A$ is an oracle and $M$ is an oracle Turing machine with oracle access to $A$, then we use the notation $M^A$.

\subsection{Reductions and Completeness}

We will define several reductions adapted from the definitions provided by Watanabe~\cite{wat:j:oneone-equivalence}. Many of these definitions can be found in earlier works. A set $A$ many-one reduces to $B$, denoted $A \leq_m B$, if there is a computable function $f$ such that $(\forall x)[x \in A \iff f(x) \in B]$.\footnote{Watanabe uses $\leq_m$ to denote polynomial-time many-one reductions, however, we have kept to the standard notation to differentiate between many-one reductions and polynomial-time many-one reductions.} This reduction is called a polynomial-time many-one reduction when $f$ is computable in polynomial time on a deterministic Turing machine and is called a linear-time many-one reduction when $f$ is computable in linear time on a deterministic Turing machine, which are denoted by $A \leq_m^{p} B$ and $A\leq_m^{linear}B$ respectively. This reduction is called a polynomial-time one-one reduction when $f$ is one-one and is computable in polynomial time, which is denoted by $A \leq_1^p B$. 
A set $A$ is length-increasing polynomial-time one-one reducible to $B$ if $A$ is polynomial-time one-one reducible to $B$ via a length-increasing function, denoted as $A \leq_1^{li} B$. A function $f$ is length-increasing if $(\forall x)[|f(x)| > |x|]$.

A language $B$ is $\exptime$-complete when $B\in\exptime$ and for all $A\in\exptime$, $A\leq_m^pB$. A language $B$ is $\etime$-complete when $B\in\etime$ and for all $A\in\etime$, $A\leq_m^{linear}B$. Finally, A language $B$ is $\pspace$-complete when $B\in\pspace$ and for all $A\in\pspace$, $A\leq_m^pB$.

\section{On Section 1 of \cite{cze:t:pspaceexp}}\label{s:sect1}
In the first section of Czerwinski's paper, four sets are defined~\cite{cze:t:pspaceexp},
\begin{align*}
&\Eset:=\{(M,x,k) \condition \text{$k$ is encoded in binary and TM $M$ accepts $x$ within $k$ steps}\},\\
&\Pset:=\{(M,x,1^k) \condition \text{TM $M$ accepts $x$ using at most $k$ tape cells}\},\\
&\Enot:=\{(M,k) \condition \text{$k$ is encoded in binary and TM $M$ accepts $\epsilon$ within $k$ steps}\}, \text{ and}\\
&\Pnot:=\{(M,1^k) \condition \text{TM $M$ accepts $\epsilon$ using at most $k$ tape cells}\},
\end{align*}
with the last two stated to be examples of $\exptime$-complete and $\pspace$-complete sets respectively. It should also be mentioned that one can easily prove that $\Eset$ and $\Pset$ are also respectively $\exptime$-complete and $\pspace$-complete. Following the definitions is Corollary~\ref{c:1}
\begin{corollary}[\cite{cze:t:pspaceexp}]\label{c:1}
    $\Eset \leq^p_1 \Enot$ and $\Pset \leq^p_1 \Pnot$.
\end{corollary}
Although this is correct, 
in later proofs there are no references made to this corollary. In addition, it is also difficult to see the importance of the sets $\Eset$ and $\Pset$, with $\Enot$ and $\Pnot$ already introduced as complete sets for their respective classes. 

We note in passing that the author introduces a lemma, whose statement is true. Although the given proof contains minor errors, we omit this discussion as the next sections' discussion is of greater importance.

After the lemma, a remark regarding the relativization barrier is given. The content of the remark offers little to no insight into how the paper plans to deal with the relativization barrier. Consequently, we assume that its sole purpose is to assure readers that the paper will take the possible issue with the relativization barrier into consideration later on in Section~5 of the paper.

\section{On Section 2 of \cite{cze:t:pspaceexp}}\label{s:sect2}
Section~2 of Czerwinski's paper focuses on undecidability when considering properties of time and space complexity. First, two notations are introduced, $s_M(x)$ and $t_M(x)$, 
which we have defined in Section~\ref{s:prelims} of our critique. Afterwards, two inequalities concerning $s_M(x)$ and $t_M(x)$ are presented, which are taken from a paper by Hopcroft, Paul, and Valiant~\cite{hop-pau-val:a:timevsspace}. For an arbitrary single-tape deterministic Turing machine $M$ and an arbitrary input $x$ where $M$ halts on $x$, it holds that:
\begin{align*}
    s_M(x) &\leq t_M(x)\text{ and}\\
    t_M(x) &= s_M(x) \times 2^{O(s_M(x))} \text{.}
\end{align*}
Nonetheless, it is unclear how this result is applied to the arguments made in this paper, as it does not seem to be used by any of the proofs.

The paper then introduces its first theorem.
\begin{theorem}[\cite{cze:t:pspaceexp}]\label{t:1}
    For an arbitrary TM $M$, any nontrivial property $\mathbf{P}$ of the functions $s_M$ and $t_M$ is undecidable. 
\end{theorem}
The theorem seems to be a modification of the well-known Rice's~theorem~\cite{ric:j:undecidable}. Instead of using nontrivial semantic properties like in Rice's~theorem~\cite{ric:j:undecidable}, the theorem considers nontrivial properties of the functions $s_M$ and $t_M$, which are the space and time complexity of $M$ on some input $x$. 

We present a counterexample of Theorem~\ref{t:1}. Let $\mathbf{P_1}(s_M,t_M)=\{M\condition$ $M$ is a Turing machine and $s_M(\epsilon)\geq 2$ and $t_M(\epsilon)\geq 2\}$, and $L_{st2} = \{M \condition M\in\mathbf{P_1}(s_M,t_M)\}$. It is easy to see that this property is nontrivial: There exists some Turing machines that satisfy $\mathbf{P_1}$, which implies $L_{st2} \neq \emptyset$, and there exists a Turing machine $M_{s1}$ that only uses one tape space on some input $x_{s1}$ and a Turing machine $M_{t1}$ that only runs for one step before halting on some input $x_{t1}$, which implies $M_{s1},M_{t1}$ is not in $\mathbf{P_1}(s_M,t_M)$. We will describe the decider $S$ for $L_{st2}$ as follows.

\begin{algorithm}[H]
\caption{Decider $S$ for $L_{st2}$}
\begin{algorithmic}
\Decider{$S(M)$:}{}
\State Simulate $M$ on $\epsilon$
\State $m\gets$ size of the tape alphabet of $M$
\If {$M$ halts on $\epsilon$ after the first step} rejects
\Else {} from the second step to the $m$th step: 
\If {$M$ reaches the second tape cell} accept
\ElsIf {$M$ halts on $\epsilon$} reject 
\EndIf
\EndIf
\State After simulating the $m$th step, reject
\EndDecider
\end{algorithmic}
\end{algorithm}

The reason why $S$ can accept after running the $m$th step is because if $M$ only uses the first tape cell and there are $m$ different tape characters, then there are only $m$ different configurations of space. Consequently, if $M$ runs more than $m$ steps using only the first tape cell, then we can be certain that $M$ is looping forever within that cell. Apart from this, it should be easy to see how $S$ always halt and $L(S) = L_{st2}$.

The proof of Theorem~\ref{t:1} has a faulty assumption that for any TM $M$ and any nontrivial property $\mathbf{P}$ of the functions $s_M$ and $t_M$, and on any input $x$, the exact value of $s_M(x)$ and $t_M(x)$ are needed to check property $\mathbf{P}$. However, in the counterexample that we created, calculating neither the exact space complexity nor the exact time complexity was necessary in deciding whether a Turing machine $M$ is in $L_{st2}$.

The paper then uses the previous result of Theorem~\ref{t:1} and extends it to Corollary~\ref{c:2}.
\begin{corollary}[\cite{cze:t:pspaceexp}]\label{c:2}
    Let $f:\naturalnumber\rightarrow\naturalnumber$ be a function with
    \[ \lim_{|x|\to\infty} \frac{f(|x|)}{t_M(x)} = 0 \] and \[ \lim_{|x|\to\infty} \frac{f(|x|)}{\log(t_M(x))} > 0. \] It is undecidable if $(\forall x) [s_M(x) \leq f(|x|)].$
\end{corollary}
Unfortunately, as we have already shown that Theorem~\ref{t:1} is flawed, the proof of Corollary~\ref{c:2} is incomplete, thus its correctness is not guaranteed.

We continue to the findings of the next and final part of this section, Corollary~\ref{c:3}.
\begin{corollary}[\cite{cze:t:pspaceexp}]\label{c:3}
    If one wants to show that an arbitrary TM $M$ does not accept an input $x$ within $t$ steps, a computational cost of $\Omega(t)$ is needed in the worst case.
\end{corollary}
While reading the proof of the corollary, we notice that the proof does not address the correctness of Corollary~\ref{c:3}. Instead, the proof tries to set up another nontrivial property of the function $t_M$ for an arbitrary Turing machine $M$, and, based on Theorem~\ref{t:1}, concludes that it is undecidable. It should also be recognized that Corollary~\ref{c:3} is in fact correct. Briefly speaking, the reason why checking whether a Turing machine $M$ accepts an input $x$ within $t$ steps might require a runtime of $\Omega(t)$ is because the overhead from simulating $M$ on $x$ could raise the total runtime to be bigger than $t$. 

\section{On Section 3 of \cite{cze:t:pspaceexp}}\label{s:sect3}

We will now critique Section~3 of Czerwinski's paper. In this section, 
the paper presents findings related to length-increasing functions and reductions.

Czerwinski's paper defines length-increasing functions similarly to Section~\ref{s:prelims} of our critique, with one additional requirement of having a polynomial-time computable inverse. 

The paper next introduces Theorem~\ref{t:2}, which is adapted from one of the results of Watanabe's paper ``On one-one polynomial time equivalence relations''~\cite{wat:j:oneone-equivalence}. 
\begin{theorem}[\cite{cze:t:pspaceexp}]\label{t:2}
    If $A$ and $B$ are $\exptime$-complete, then $A \leq_1^{li} B$ and $B \leq_1^{li} A$ hold.
\end{theorem}

As a proof for this theorem, Czerwinski cites the proof of Corollary~3.4 from Watanabe's paper~\cite{wat:j:oneone-equivalence}. It should be noted that at the time of Watanabe's paper, it was standard to define $\exptime$ to represent what we have defined as $\etime$. As a result, Watanabe's paper only provides a proof where $A$ and $B$ are $\etime$-complete, rather than being $\exptime$-complete. Czerwinski's paper does not provide a preliminaries section, so it is unclear if the paper intends to use $\etime$ or $\exptime$. However, we believe its main goal is to separate $\exptime$ and $\pspace$, as the result $\etime \neq \pspace$ is already well-known~\cite{boo:j:comparing-complexity-classes}. Moreover, Watanabe~\cite{wat:j:oneone-equivalence} does not define length-increasing functions to be necessarily polynomial-time invertible, which is different from the definition given in Czerwinski's paper.
For these reasons, 
the proof of Theorem~\ref{t:2} is invalid. 
This is a critical flaw in Czerwinksi's paper since the result of Theorem~\ref{t:2} sets the stage for Theorem~\ref{t:3} to be able to show that $\pspace \neq \exptime$. We will discuss this more in Section~\ref{s:sect4}.   

\setcounter{lemma}{1}
Next, we will critique Lemma~\ref{l:2}.  
\begin{lemma}[\cite{cze:t:pspaceexp}]\label{l:2} 
    Let $A, B \subseteq \{(M, k) \condition \text{TM $M$ accepts $k$}\}$ and $A \leq_1^{li} B$. Then there is a length-increasing reduction\footnote{Originally in Czerwinski's paper, $g$ is stated to be a length-increasing function. However, if $g$ is simply a length-increasing function, then the if and only if condition that follows is flawed in that it mishandles cases of strings that are syntactically not of the form $(\cdot,\cdot)$. Therefore, we suspect what the author intends is for $g$ to be a length-increasing reduction instead. Since this does not significantly affect the critique, we leave it as a footnote.} $g$ such that:
    $$(M, k) \in A \iff g(M, k) \in B,\text{ and}$$
    $$g(M,k)=(M_g, k_g)$$
    where $M_g$ depends only on $M$. 
\end{lemma}

Before going into the proof, we would like to introduce some notations that the author uses. Functions $g_1$ and $g_2$ are defined as $g_1(M,k)=M_g$ and $g_2(M,k)=k_g$, and thus $(M_g,k_g)=g(M,k)=(g_1(M,k),g_2(M,k))$. By the S-m-n theorem, $g_1(M,k)=g_{1,S(M)}(k)$ and $g_2(M,k)=g_{2,S(M)}(k)$. Additionally, by Czerwinski's definition of a length increasing function, $g$ is polynomial-time invertible, and thus $(M,k)=(g_{1,S(M)}^{-1}(k_g),g_{2,S(M)}^{-1}(k_g))$.

We would like to focus on the pseudocode given at the end of the proof.
\begin{algorithm}[H]
\caption{}
\begin{algorithmic}
\Def $M_g(k_g)$:
\State $\hat{k} \gets g^{-1}_{2,S(M)}(k_g)$
\If {$\hat{k}$ is defined}
\State \Return $g_{1,S(M)}(\hat{k})(k_g)$
\EndIf
\EndDef
\end{algorithmic}
\end{algorithm}

This pseudocode is provided as an algorithm in which $M_g$ can be constructed and simulated on $k_g$ without using $k$, thus making it depend only on $M$. To our understanding, given $k_g$ as some input, the algorithm will construct $\hat{k}$ that, if defined, satisfies that $(M,\hat{k}) \in A$. Afterwards, $M_g$ can be constructed with $g_{1,S(M)}(\hat{k})$. However, this construction does not prove the statement of the lemma. The lemma claims that $M_g$ only depends on $M$, while the provided pseudocode also uses $k_g$ in its calculation.

\section{On Section 4 of \cite{cze:t:pspaceexp}}\label{s:sect4}
The main result is presented in Section~4 of Czerwinski's paper.
\begin{theorem}[\cite{cze:t:pspaceexp}]\label{t:3}
    There is no $\leq^{li}_1$-reduction from $\Enot$ to $\Pnot$.
\end{theorem}

It is crucial to see that in order for this theorem to imply that $\exptime \neq \pspace$, then Theorem~\ref{t:2} must hold. 
Theorem~\ref{t:2} proposes that all $\exptime$-complete sets are length-increasing polynomial-time one-one reducible to each other. Hypothetically, if Theorem~\ref{t:2} was in fact true, then one would only need to show that there is no length-increasing polynomial-time one-one reduction from a $\exptime$-complete set, $\Enot$, to a $\pspace$-complete set, $\Pnot$, to prove that $\Pnot$ is not $\exptime$-complete and thus $\exptime\neq\pspace$, which is the statement of Theorem~\ref{t:3}.
As we have identified the problems with Theorem~\ref{t:2} earlier, if Theorem~\ref{t:3} were to hold, it would not be strong enough to imply that $\pspace \neq \exptime$. 
Additionally, the proof of this theorem also uses Lemma~\ref{l:2} and Corollary~\ref{c:2}, which we have shown to be invalid. As a result, we can say that this proof is similarly incomplete, and ultimately, the paper fails to show that $\pspace \neq \exptime$.

\section{On Section 5 of \cite{cze:t:pspaceexp}}\label{s:sect5}
In this section, Czerwinski attempts to argue that the proof technique does not violate the relativization barrier. In order to do this, he introduces Proposition~\ref{p:1}. 

\begin{proposition}[\cite{cze:t:pspaceexp}]\label{p:1}
There is a computable function $f$ such that $(\forall (M,x,k))[(M, x, k) \in \Eset \iff f((M, x, k)) \in \Pset]$ 
with exponential computation time.
\end{proposition}

As a proof for this proposition, the paper provides pseudocode that takes a problem from $\Eset$ and computes a set in $\Pset$ in exponential time. The pseudocode seems correct, however the proposition itself is trivial.
It is unclear how this proposition demonstrates that the purported proof technique does not violate the relativization barrier.  

The paper claims that the existence of this function $f$ being a many-one reduction that is computable but not computable in polynomial-time ``means $\Eset \leq_m \Pset$, but $\Eset \not\leq_m^p \Pset$ and so, $\pspace^{\p} \neq \exptime^{\p}$"~\cite{cze:t:pspaceexp}. 
Prior to this point, the paper has not directly made the claim that $\Eset \not\leq_m^p \Pset$ so it is unclear how the existence of a exponential time reduction prevents a polynomial time one from existing. In fact, the existence of a polynomial-time many-one reduction would imply the existence of a exponential-time many-one reduction. 

The claim that $\Eset \not\leq_m^p \Pset$ implies that $\pspace^{\p} \neq \exptime^{\p}$ is correct, however it is unclear why the paper uses the classes with the $\p$ oracles rather than the classes directly as $\Eset \not\leq_m^p \Pset$ more directly implies that $\pspace \neq \exptime$ which then implies $\pspace^{\p} \neq \exptime^{\p}$ since $\pspace^{\p} = \pspace$ and $\exptime^{\p} = \exptime$. 

\section{Conclusion}\label{s:conclusion}
In this critique, we have pointed out errors in ``Separation of $\pspace$ and $\exptime$''~\cite{cze:t:pspaceexp} and come to the conclusion that the paper fails to show that $\pspace \neq \exptime$. The paper makes several mistakes in its reasoning and 
thus is unable to provide a sufficient proof for its key claim.
As a result, it is still an open issue whether $\pspace \neq \exptime$.

\paragraph{Acknowledgements}
We would like to especially thank Michael C. Chavrimootoo for his helpfulness during the development of our critique and his helpful comments on previous drafts. We would also like to thank 
Ethan Ferland, 
Lane A. Hemaspaandra, 
and
David E. Narv\'{a}ez
for their helpful comments on prior drafts.
The authors are solely responsible for any remaining errors.

\bibliographystyle{alpha}
\bibliography{gry-reu}

\begin{thebibliography}{HPV77}

\bibitem[AB09]{aro-bar:b:complexity}
S.~Arora and B.~Barak.
\newblock {\em Complexity Theory: A Modern Approach}.
\newblock Cambridge University Press, 2009.

\bibitem[Boo74]{boo:j:comparing-complexity-classes}
R.~Book.
\newblock Comparing complexity classes.
\newblock {\em Journal of Computer and System Sciences}, 3(9):213--229, 1974.

\bibitem[Cze21]{cze:t:pspaceexp}
R.~Czerwinski.
\newblock Separation of {PSPACE} and {EXP}.
\newblock Technical Report arXiv:2104.14316~[cs.CC], Computing Research
  Repository, \mbox{arXiv.org/corr/}, April 2021.
\newblock Version~1.

\bibitem[HO02]{hem-ogi:b:companion}
L.~Hemaspaandra and M.~Ogihara.
\newblock {\em The Complexity Theory Companion}.
\newblock Springer-Verlag, 2002.

\bibitem[HPV77]{hop-pau-val:a:timevsspace}
J.~Hopcroft, W.~Paul, and L.~Valiant.
\newblock On time versus space.
\newblock {\em Journal of the ACM}, 24(2):332--337, April 1977.

\bibitem[Kle52]{kle:b:metamathematics}
S.~Kleene.
\newblock {\em Introduction to Metamathematics}.
\newblock {D. van Nostrand Company, Inc.}, 1952.

\bibitem[Ric53]{ric:j:undecidable}
H.~Rice.
\newblock Classes of recursively enumerable sets and their decision problems.
\newblock {\em Transactions of the American Mathematical Society},
  74(2):358--366, 1953.

\bibitem[Sip13]{sip:b:introduction-third-edition}
M.~Sipser.
\newblock {\em Introduction to the Theory of Computation}.
\newblock Cengage Learning, 3rd edition, 2013.

\bibitem[Wat85]{wat:j:oneone-equivalence}
O.~Watanabe.
\newblock On one-one polynomial time equivalence relations.
\newblock {\em Theoretical Computer Science}, 38:157--165, 1985.

\end{thebibliography}
	
\end{document}